# Nash Equilibria of The Multiplayer Colonel Blotto Game on Arbitrary Measure Spaces

Siddhartha Visveswara Jayanti[*]

సిద్ధార్థ విశ్వేశ్వర జయంతి

सिद्धार्थ विश्वेश्वर जयन्ति

Massachusetts Institute of Technology
Computer Science and Artificial Intelligence Laboratory (CSAIL)

January 2021

**Abstract**

The Colonel Blotto Problem proposed by Borel in 1921 has served as a widely applicable model of budget-constrained simultaneous winner-take-all competitions in the social sciences. Applications include elections, advertising, R&D and more. However, the classic Blotto problem and variants limit the study to competitions over a finite set of discrete battlefields. In this paper, we extend the classical theory to study multiplayer Blotto games over arbitrary measurable battlegrounds, provide an algorithm to efficiently sample equilibria of symmetric "equipartionable" Generalized Blotto games, and characterize the symmetric fair equilibria of the Blotto game over the unit interval.

## 1 Introduction

The *Colonel Blotto game*, proposed by Borel in 1921 [11], has been a staple of the game theory literature for the past century. In this game, two warring commanders, Captain Alice and Colonel Blotto, fight on $n$ simultaneous and independent *battlefields* of different values $v_1, \ldots, v_n$. Each commander has a finite budget of *power*—$\mathcal{B}_{Alice}, \mathcal{B}_{Blotto}$—to split over the battlefields, and for each battlefield $i$, the value $v_i$ will be awarded to the commander that allocates more power there. Each commander aims to maximize the total expected value of his/her winnings in this game. The game is called *symmetric* if all players have the same budget, *homogeneous* if all battlefields have the same value, *continuous* if allocations can be arbitrary positive real numbers, and *discrete* if allocations are restricted to be non-negative integers. In this paper, we define a version of the Blotto game that not only encompasses all previously studied variants of the game, but also extends the scope of the game to settings where the competition is not over a finite set of discrete battlefields 1,...,$n$, but over arbitrary measurable battlegrounds such as intervals of the real line, the surface of the earth, etc. Such Blotto games can model social competitions ranging from nations fighting wars over a terrain, to mobile data providers fighting over customer bases in a country. In this paper, we focus on computing and characterizing the *Nash Equilibria*[1] of such *Measure Space Colonel Blotto games*.

---

[*]Funded by the Department of Defence through the NDSEG fellowship.

The Blotto game has proved to be a treasure trove for mathematicians, computer scientists, and social scientists. While the Blotto game is easy to describe, its equilibrium structure is surprisingly complex even in the mathematically pleasing continuous setting, due to the hard budget constraint.[2] Thus, an analysis of the equilibria of even the two-player game has only emerged in the past few decades due to a long line of research [24, 37, 31, 21, 33, 35] which has culminated in the work of Kovenock and Roberson [22]. Roberson's complete characterization of the equilibria of homogeneous two-player Blotto is a highlight of this line of work [31].

When the allocations must be discrete, closed form analyses have given way to fast algorithms that can output equilibria given a problem description; thus, computer scientists have taken the fore, producing several pseudo-polynomial time algorithms for various settings of the discrete Blotto problem [1, 7, 5, 6]. The Blotto game has also seen innumerable applications in the social sciences, ranging from modeling political elections [27, 25, 24, 26, 10] to modeling competitions in research and development [17, 23].

The most famous application of the Blotto game is in the analysis of elections. Here the two commanders represent two parties, the battlefields represent voters, and the allocation describes the amount of public resources that each voter will benefit from due to the party's platform [27, 25, 24, 26, 31, 22, 10]. For this reason, we call a player's (party's) randomized allocation strategy *fair*, if its allocation to each battlefield (voter) is identical (in distribution). Furthermore, since each voter has a single vote, the homogeneous setting is of particular interest, and since the pool of allocatable public resources is independent of party, most of the work on Blotto focuses on symmetric games [24, 25, 26, 27, 10].

Equilibria for the (hard budget constraint) Blotto games have been developed based on solutions for the simpler soft budget constraint version of the game, called *General Lotto*. In the Lotto game, each player plays a *distribution* over allocations instead of a single allocation, and the budget constraint is relaxed so that the allocations need to sum to the budget *only in expectation*, rather than with probability one. Strategies in the Lotto game are a superset of mixed strategies in the Blotto game, since Lotto players can play strategies that go over-budget sometimes as long as they compensate by playing strategies that go under-budget some other times.[3] It turns out that this freedom usually makes Lotto easier to solve in practice than Blotto. In particular, the distributions over strategies for each battlefield can be computed independently for Lotto games, as long as the expected values of the resultant distributions played on each battlefield add up to the budget constraint. In contrast, a mixed-strategy for Blotto with the same marginal allocation distributions per battlefield would additionally need to couple these marginals into a joint distribution whose support only consists of allocation vectors summing to the budget constraint. Demonstration of such a coupling is often the hardest part of an equilibrium computation for a Blotto game.

Myerson was the first to use a Blotto-like game to model multiparty elections, such as the parliamentary elections of India or the United Kingdom, in [27], which compares different election systems by studying the equilibrium strategies those systems induce. However, stating that "the hardest part of [the Blotto] problem was to construct joint distributions for allocations that always sum to the given total," Myerson weakened the true budget constraint to the soft one and characterized the equilibria of multiplayer homogeneous symmetric General Lotto. It took another three decades before Boix-Adserà et al. showed many settings under which Myerson's marginals could be coupled into budget-respecting joint distributions, thereby finding equilibria of the multiplayer Blotto game [10]. In particular, they designed an efficient algorithm that samples Nash equilibrium strategies for most symmetric three-player Blotto games, and homogeneous symmetric Blotto games with $k > 3$ players given that the number of players divides the number of battlefields. Recently, Naskar et al. have introduced a quantum version of the game, and demonstrated that players with access to quantum strategies have an advantage over those with classical strategies [30, 29].

---

[1] A collection of prescribed strategies, one for each player, is in *Nash equilibrium* if no player can increase his expected winnings by unilaterally deviating from the prescription. If the prescribed strategies are deterministic, the equilibrium is *pure*; if they are stochastic, the equilibrium is *mixed*.

[2] Even the simplest Blotto games do not admit pure equilibria. Consider the two-player continuous symmetric homogeneous Blotto game with $n > 2$ battlefields. If Alice plays any deterministic strategy, Blotto can choose to allocate nothing to one of the battlefields and win each of the remaining battlefields by playing $\epsilon$ more than her there.

[3] For example, if Alice has a budget of $\mathcal{B}_{Alice} = 1$ she can play $\vec{a} = (1,1)$ with probability $1/2$ and $\vec{a'} = (0,0)$ with probability $1/2$ in the Lotto game; but cannot in the Blotto game because, $||\vec{a}||_1 = 1 + 1 = 2 > 1 = \mathcal{B}_{Alice}$.

In the century after its introduction by Borel, the Blotto game has seen a multitude of applications in modeling: Elections, where the players are parties and the battlefields are voters (or voting districts) [27, 25, 24, 26, 10]; R&D competitions, where the players are companies and battlefields are product development projects [17, 23]; advertising, where the players are companies advertising substitute goods and the battlefields are consumers [16]; and even ecology, where the players are species competing for habitats and the battlefields are the ecological niches that they are hoping to fill [17]. Additionally, there are mathematical connections between the Blotto game and simultaneous all-pay auctions [31, 32].

The wide success of Blotto in modeling two-player and multiplayer social competitions over discrete sets of $n$ items begs the natural question: what about competitions over other sets, such as ecological competitions over the area of a forest or ideological battles over the surface of the earth? This paper provides the mathematical underpinnings to model such problems by extending the definition of the Blotto problem to arbitrary measure spaces, and providing a Nash equilibrium analysis of a class of such Measure Space Blotto games.

## 1.1 Informal Description of Measure Space Blotto

In our *Measure Space Blotto* game, $k$ players compete over a measurable *battleground* $\Omega$, which is equipped with two measures: the budget measure $\beta$ and the value measure $\upsilon$. It is illustrative to imagine $\Omega$ to be the unit-interval $[0, 1] \subseteq \mathbb{R}$ with both $\beta$ and $\upsilon$ equalling the standard Lebesgue measure $\lambda$ on the interval. Each player $i$ plays a function $\phi_i : \Omega \to \mathbb{R}_+$, and wins the subset of points $W_i \subseteq \Omega$ where his function $\phi_i$ is larger than every other function $\phi_j$. Player $i$'s objective is to maximize the value, $\upsilon(W_i)$, of his winnings, while satisfying the budget constraint, i.e., ensuring that the integral $\int_\Omega \phi_i d\beta$ is at most $\mathcal{B}_i$. Thus, intuitively, the budget measure specifies how difficult it is to compete in a certain part of the battleground, while the value measure describes how valuable that part of the battleground is. The motivating examples below show why these two measures may be quite different, while also demonstrating the applicability of the problem.[4] Our Measure Space Blotto problem subsumes the standard multiplayer Blotto problem; to see this, just set $\Omega = \{1, ..., n\}$, $\beta$ to the counting measure, and $\upsilon$ to value the $j$th battlefield at $v_j$.

## 1.2 Motivations for Measure Space Blotto

The Measure Space Blotto game that we introduce in this paper has potential applications to many exciting areas of study in the social sciences including:

---

[4]The informal description suffices to understand the introduction. Section 3 contains the formal definition.

- **War of Ideologies:** $k$ different ideologies compete over the surface of the earth. The budget measure is determined by how difficult it is to compete in a given region of the earth, while value is assigned by how important a given piece of land is (by population density, for instance).

- **Data Providers:** $k$ mobile data plans compete over the surface of the earth. Consumers in a given place will opt for the data plan that has better signal and service in that place. The cost of providing a given signal strength varies with the terrain and accessibility across the world, and the value of being the best provider increases with population density.

- **Colonial Powers** $k$ different colonial powers compete over a region of the world that they intend to conquer. The budget measure is determined by how difficult a particular region is to conquer, while value is determined by the worth of natural resources that the winning country will be able to control.

- **Standard Multiplayer Blotto** The Measure Space Blotto problem subsumes the standard Multiplayer Blotto problem: just set $\Omega = \{1, ..., n\}$, $\beta$ to the counting measure, and $v$ to value the $j$th battlefield at $v_j$.

In addition to the potential applications in the social sciences, Measure Space Blotto also captures problems of mathematical interest such as:

- **Interval Blotto:** $k$ symmetric players, each with budget one, compete over the compact interval $[0, 1]$. Each Player $i$ can play a measurable function $\phi_i : [0, 1] \to \mathbb{R}_+$, and gets utility equal to the measure of the subset of points where his function is the largest. How will the players play this game?

- **Blotto on the Sphere:** Identical to Interval Blotto, but played over the surface of the standard sphere $\mathbb{S}_2$, rather than over the compact interval $[0, 1]$.

## 1.3 Our Contributions

We make three contributions in this paper:

1. We introduce the novel *Measure Space Blotto* and *Lotto* games which model multiplayer budget-constrained competitions over arbitrary measurable battlegrounds.

2. We extend Boix-Adserà et al. [10]'s theory for multiplayer standard Blotto to solve for equilibria in a large class of Measure Space Blotto games. In particular, we compute mixed Nash equilibria for the class of symmetric "equipartitionable" Measure Space Blotto games for any number of players $k \in [1, \infty)$. The class includes Interval Blotto, Blotto on the Sphere, as well as the case of multiplayer Blotto that has been solved for $k > 3$ players. Furthermore, We design an efficient, $O(k)$ time algorithm for each player to sample an equilibrium strategy.

3. We characterize the symmetric fair equilibria of Interval Blotto for an arbitrary number of players. We believe the characterization is the technical highlight of our paper.

    To the best of our knowledge, Roberson's equilibrium characterization of homogeneous two-player Blotto is the only previously known characterization for a hard-constraint Blotto game. Roberson's approach exploited the fact that there are a finite set of $n$ battlefields to set up a Lagrangian system, which explicitly involved $2n$ CDFs—one for each marginal distribution

that each player plays on each battlefield. He then used techniques from all-pay auction analyses to obtain his result [31].

Our Interval Blotto setting has a continuum of points in the battleground, thereby ruling out Roberson's methods which relied on their being a finite number of points in the battleground set. So, we develop a different approach to characterizing Blotto equilibria. Our method exploits the continuity of the battleground to create a mapping from marginal allocation distributions on individual points of the battleground to strategies in the game. Curiously, our mapping allows us to leverage some tools from Myerson's work on characterizing equilibria of standard $n$ battlefield Lotto that could not be adapted to standard Blotto—in fact, Myerson states: "the advantage of my simplified formulation is that it will enable us to go beyond this 'Colonel Blotto' literature and get results about more complicated situations in which more than two candidates are competing." to justify analyzing Lotto instead of Blotto [27]. Ultimately, our method allows us to directly characterize equilibria of Interval Blotto for any number of players without going through any external mechanism like all-pay auctions.

## 1.4 Organization of the Paper

The remainder of this paper contains five sections. We describe prior work in Section 2, and define the Measure Space Lotto and Blotto games in Section 3. We solve for Nash equilibria of equiparitionable symmetric Measure Space Blotto games, and give efficient algorithms to sample our equilibria in Section 4. We build on the equilibrium solution in the previous section to characterize the symmetric fair equilibria of Interval Blotto in Section 5. We end with some remarks in Section 6.

## 2 Prior work

The Colonel Blotto game was introduced by Émile Borel in 1921, exactly a century ago [11]—almost three decades prior to Nash's seminal work on equilibria of games [28]. The abundance of social scientific applications that are modeled by Blotto [26, 3, 13, 16, 4, 14] (detailed in the introduction), have made analyzing the equilibria of the Blotto game a hotbed of research ever since.

Initially, progress on the problem was slow: it took 17 years for Borel and Ville to discover the equilibria of the symmetric homogeneous game with just three battlefields [12], and another dozen years for Gross and Wagner to extend the solutions to the symmetric heterogeneous case with $n$ battlefields [19, 18]. Published progress on the problem was subsequently hindered by secrecy concerns during the second world war [9].

Blotto's applications to understanding electoral politics brought it back into vogue at the turn of the millennium [25, 24]. This resurgence was punctuated by Roberson's landmark thesis, where he characterized equilibria for all two-player $n$ battlefield homogeneous Blotto games [31]. Roberson's result was the first to tackle any part of the erstwhile inscrutable non-symmetric setting of the game, and his new insights into the problem spurred a series of papers inching ever closer to solving the elusive general case of Blotto [33, 22, 35]. Kovenock and Roberson's 2015 paper summarizes the state-of-the-art in two-player Blotto [22].

A parallel stream of work on General Lotto was set-off by Bell and Cover in 1981 [8]. Myerson linked the two streams of research when he modeled multiparty elections using the multiplayer symmetric homogeneous Lotto game, and characterized the fair equilibria of such games [27]. Recently, Boix-Adserà et al. introduced the multiplayer Blotto game in 2020 [10]. They expanded Myerson's Lotto equilibrium solutions to the heterogeneous setting, showed how to couple the marginal allocation distributions into a joint distribution Blotto equilibrium under mild conditions, and derived efficient algorithms that players could use to sample equilibrium strategies. More recently, Naskar and Maioli considered a quantum version of multiplayer Blotto, and showed that quantum strategies can outperform classical ones [30, 29].

The discrete Blotto game has been studied vigorously in the past dozen years. Algorithms for computing equilibria of discrete symmetric Blotto were given by Hart [20], and discrete non-symmetric Lotto by Dziubinski et al. [15]. And a host of recent works computed solutions for variants of Blotto games [21, 1, 7, 5, 6, 36, 34, 23, 17].

## 3 The Measure Space Blotto Game

In the standard multiplayer Colonel Blotto game, players compete over a finite discrete set of $n$ battlefields. Motivated by a desire to model the vast array of competitions described in Section 1.2, we define a generalized version of the game, where players compete over arbitrary measure spaces. We have already given an informal explanation of the game in Section 1.1; we formalize it below.

**Definition 3.1.** An instance of the *Measure Space Blotto Game* is defined by the six-tuple:

$$\left(k, \Omega, \mathbb{B}, \vec{\mathcal{B}}, \beta, \upsilon\right)$$

- $k \in \mathbb{N}$ is the number of players.
- $\Omega$ is a measure-space called the *battleground* which is the domain of each player's bid function.
- $\mathbb{B} \subseteq \mathbb{R}_+$ is the co-domain of each player's bid function.
- $\vec{\mathcal{B}} \in \mathbb{R}_+^k$ is the tuple of budget constraints. Specifically, Player $i$ has a budget of $\mathcal{B}_i$.
- $\beta$ is a finite measure on $\Omega$, called the *budget measure*.
- $\upsilon$ is a finite on $\Omega$ called the *value measure*. We denote the total value of the battleground to be $\Upsilon \triangleq \upsilon(\Omega)$, and impose that this be a bounded quantity. Furthermore, we impose that $\beta$ and $\upsilon$ are absolutely continuous with respect to each other, so that $\frac{d\beta}{d\upsilon}$ and $\frac{d\upsilon}{d\beta}$ are both defined by the Radon–Nikodym theorem.
- We call the game *symmetric* if $\vec{\mathcal{B}} = \vec{\mathbb{1}}$, and *homogeneous* if both $\beta$ and $\upsilon$ are the uniform measure on the space $\Omega$.

Each Player $i \in [k]$ plays a bounded measurable *bid function* $\phi_i : \Omega \to \mathbb{B}$ satisfying the budget constraint

$$\int_\Omega \phi_i(x) d\beta(x) \leq \mathcal{B}_i.$$

For a given *bid profile* $\phi = (\phi_i)_{i \in [k]}$, the utility of Player $i$ is given by the expression

$$U_i(\phi) = \int_\Omega \frac{\mathbb{1}\left(i \in \arg\max_{j \in [k]}\{\phi_j(x)\}\right)}{\left|\arg\max_{j \in [k]}\{\phi_j(x)\}\right|} \, d\upsilon(x)$$

Often, we will need to talk about mixed strategies for the Measure Space Blotto game. A *mixed strategy* of Player $i$ is a distribution $\Phi_i$ (over functions $\phi_i$). We call such a strategy $\Phi_i$ *fair* if $\Phi_i(x)$ is distributed identically for (almost) every point $x \in \Omega$. Once each Player $i$ has fixed a mixed strategy $\Phi_i$, we call the tuple of strategies $\Phi = (\Phi_i)_{i \in [k]} = (\Phi_1, \Phi_2, \ldots, \Phi_k)$, the *strategy profile*. where each $\Phi_i$ is a distribution over valid functions $\phi_i$. △

**Remark 3.2.** *The measure space $\Omega$ technically consists of the set $\Omega$ along with a $\sigma$-algebra $\mathcal{A}$. For the purposes of this paper, it suffices to leave the $\sigma$-algebra implicit—we use the standard Lebesgue $\sigma$-algebra when dealing with Euclidean spaces such as the interval $[0, 1]$ in Interval Blotto.*

A mixed strategy for Player $i$ in a Blotto game is some distribution-over-functions $\Phi_i$, such that *every realization $\phi_i \in \text{supp}(\Phi_i)$ satisfies the budget constraint*. This hard-budget constraint makes it notoriously difficult to solve for Nash equilibria in Colonel Blotto games [27]. So, researchers who have analyzed Colonel Blotto games have found it convenient to proceed by first analyzing its cousin General Lotto game which imposes a weaker soft-budget constraint on mixed strategies, namely, that the budget constraint need only be solved in expectation. Even in our measure space context, we can define such a cousin for the Measure Space Blotto game, which we call *Measure Space Lotto*.

**Definition 3.3.** An instance of *Measure Space Lotto* is specified by the same tuple $\left(k, \Omega, \mathbb{B}, \vec{\mathcal{B}}, \beta, \upsilon\right)$. In a game of Lotto, each Player $i$ plays a distribution over functions, $\Phi_i$, that satisfies the property

$$\mathbb{E}_{\phi_i \sim \Phi_i} \left[ \int_\Omega \phi_i(x) d\beta(x) \right] \leq \mathcal{B}_i.$$

Given a strategy profile $\Phi = (\Phi_i)_{i \in [k]}$ of the $k$ players, fixed functions $(\phi_i \sim \Phi_i)_{i \in [k]}$ for each player are sampled independently from the distribution over functions and each Player $i$'s pay off is calculated as

$$U_i(\phi) = \int_\Omega \frac{\mathbb{1}\left(i \in \arg\max_{j \in [k]}\{\phi_j(x)\}\right)}{\left|\arg\max_{j \in [k]}\{\phi_j(x)\}\right|} d\upsilon(x).$$

And thus the expected utility for Player $i$ is $U_i(\Phi) = \mathbb{E}_{\phi \sim \Phi}(U_i(\phi))$. △

As a warm-up and sanity check, we show that Measure Space Blotto is indeed a constant sum multiplayer game.

**Lemma 3.4.** *For any bid profile $\phi = (\phi_1, \ldots, \phi_k)$ played in a $k$-player Measure Space Blotto game $\left(k \in \mathbb{N}, \Omega, \vec{\mathcal{B}} \in \mathbb{R}_+^n, \beta, \upsilon\right)$, the total sum of the utilities is the total value of the battleground, i.e.,*

$$\sum_{i=1}^{k} U_i(\phi) = \Upsilon = \upsilon(\Omega)$$

*Proof.* We prove the lemma via the linearity of expectation.

$$\sum_{i=1}^{k} U_i(\phi) = \sum_{i=1}^{k} \int_\Omega \frac{\mathbb{1}\left(i \in \arg\max_{j \in [k]}\{\phi_j(x)\}\right)}{\left|\arg\max_{j \in [k]}\{\phi_j(x)\}\right|} \, dv(x)$$

$$= \int_\Omega \frac{\sum_{i=1}^{k} \mathbb{1}\left(i \in \arg\max_{j \in [k]}\{\phi_j(x)\}\right)}{\left|\arg\max_{j \in [k]}\{\phi_j(x)\}\right|} \, dv(x)$$

$$= \int_\Omega 1 \, dv(x) = v(\Omega) = \Upsilon$$

$\square$

Notice that the proof of Lemma 3.4 does not depend on the functions $\phi$ satisfying any budget constraints, and therefore applies to Measure Space Lotto also.

## 3.1 Formal Definitions of Specific Measure Space Blotto Games

We are now ready to translate our informal descriptions of games from Section 1.2 into formal notation. In particular, we demonstrate that Measure Space Blotto captures all previous considered (finite battlefield) Blotto games by defining Multiplayer Blotto in our Measure Space Blotto formalism; then we formally define Interval Blotto, whose analysis is the focus of Section 5.

**Multiplayer Blotto:** $k$ players compete over $n$ battlefields, $[n] = \{1, \ldots, n\}$. Since the $i^{th}$ battlefield is valued at $v_i$, we set $v$ to the discrete measure $\vec{v}$, and since allocations to every battlefield contribute uniformly to the budget, $\beta$ is the counting measure, which we denote by $\#$. Thus, the game is $(k, [n], \mathbb{B}, \vec{\mathcal{B}}, \#, \vec{v})$, where $\mathbb{B} = \mathbb{R}_+$ if the game is continuous, $\mathbb{B} = \mathbb{Z}_+$ if the game is discrete, and $\mathbb{B} = \{0, 1\}$ if the game is "boolean" (as in Boix-Adserà et al. [10]).

**Interval Blotto:** $k$ symmetric players, with equal budgets ($\vec{\mathcal{B}} = \vec{\mathbb{1}}$) play non-negative functions over the space $\Omega = [0, 1]$. Both budget and value are measured uniformly, so the game is defined as $(k, [0, 1], \mathbb{R}_+, \vec{\mathbb{1}}, \lambda, \lambda)$. Understanding the equilibria of this game is the most mathematically fundamental problem in Measure Space Blotto games.

## 4 Calculating Equilibria of Symmetric Measure Space Blotto Games

For the remainder of this paper, we restrict our focus to analyzing games that are *symmetric* and *continuous*, i.e. we fix $\vec{\mathcal{B}} = \vec{\mathbb{1}}$ and $\mathbb{B} = \mathbb{R}_+$. Furthermore, in order to simplify our expressions, we normalize the finite budget measure such that $\beta(\Omega) = 1$. Our primary goal is to characterize the symmetric equilibria of Interval Blotto—the most fundamental problem in measure space Blotto games. So, we turn our attention to identifying symmetric equilibria for a large class of Measure Space Blotto games that includes Interval Blotto and Blotto on the Sphere.

Even two-player Interval Blotto has no pure Nash equilibrium: if Player 1 commits to the deterministic strategy $\phi_1 : [0, 1] \to \mathbb{R}_+$, then Player 2 can win $1 - \delta$ of the interval (for any $\delta > 0$) by playing a function $\phi_2$ equal to 0 on a $\delta$ fraction of the interval and (ever so slightly) greater than $\phi_1$ on the remaining $1 - \delta$ fraction of the interval. However, such a strategy profile cannot be in equilibrium, since Player 1 could switch his strategy from $\phi_1$ to $\phi_2$ and thereby win half of the total utility. So, in general, we are interested in solving for mixed Nash equilibria $\Phi$.

Given a distribution over functions, $\Phi_i$, we can take its restriction to a given point $x \in \Omega$ to obtain a random variable $\Phi_i(x)$. We call this random variable the *marginal* of $\Phi_i$ at the point $x$. In the next lemma, we formulate a sufficient condition for a mixed strategy profile $\Phi$ to be an equilibrium of a symmetric Blotto or Lotto game. Quite beautifully, this condition only depends upon the point-wise marginals $\Phi_i(x)$, and not on the full joint distribution of the distribution-over-functions $\Phi_i$. We will later use the freedom inherent in this sufficient condition to calculate mixed equilibria.

## 4.1 The Beta Distribution and Sufficient Conditions for Symmetric Equilibrium

It turns out that equilibria of symmetric Blotto problems are intimately tied to the Beta distribution. So, we describe the distribution here before presenting the sufficient condition.

**Definition 4.1.** For any $a, b > 0$, the Beta$(a, b)$ distribution is the distribution supported on the interval $[0,1]$ with PDF proportional to $x^{a-1}(1-x)^{b-1}$. In particular, when $b = 1$, we have Beta$(a, 1)$ is $\mathbb{P}(X \leq x) = x^a$. △

**Definition 4.2.** The *Dirichlet distribution* $\mathrm{Dir}(\alpha_1, \ldots, \alpha_m)$ is the distribution on the $(m-1)$-simplex $\Delta_{m-1}$ with density function $f(\vec{x}; \vec{\alpha}) \propto \prod_{i=1}^m x_i^{\alpha_i - 1}$, i.e., if $(X_1, \ldots, X_m) \sim \mathrm{Dir}(\alpha_1, \ldots, \alpha_m)$, then
(i) For each $i \in [m]$, $X_i \sim \mathrm{Beta}\left(\alpha_i, \sum_{j \neq i} \alpha_j\right)$.
(ii) $\sum_{i=1}^m X_i = 1$ almost surely

△

The Beta distribution's relation to the Lotto problem first appeared in Myerson's analysis of multiplayer symmetric homogeneous ($n$ battlefield) Lotto [27], where he showed that each player bidding according to a scaled Beta distribution on each of the battlefields is the unique symmetric fair equilibrium of the symmetric homogeneous Lotto game. These ideas were extended by Boix-Adserà et al. in their analysis of multiplayer symmetric Blotto [10]. In particular, they showed that even heterogeneous Lotto games have mixed equilibria with scaled Beta marginals. Furthermore, they algorithmically constructed budget-respecting joint distributions that couple these marginals when either: (1) there are exactly $k = 3$ players, and no battlefield is valued at more than $1/3$ of the total value of all battlefields, or (2) there are $k > 3$ players and the battlefields can be partitioned into $k$ sets of equal value.

In this section, we extend Boix-Adserà et al.'s techniques to derive efficiently sampleable equilibrium strategies for symmetric games over "equipartitionable" measure-spaces (defined after the next lemma). Our solutions start with the following sufficient condition.

**Lemma 4.3.** *Let $\Phi = (\Phi_i)_{i \in [k]}$ be a valid mixed strategy profile for the k-player Measure Space Blotto or Lotto game $\left(k, \Omega, \mathbb{R}_+, \vec{\mathbb{1}}, \beta, \upsilon\right)$. If*

$$\Phi_i(x) \sim \frac{k}{\Upsilon} \cdot \frac{d\upsilon}{d\beta}(x) \cdot \mathrm{Beta}\left(\frac{1}{k-1}, 1\right)$$

*for every $i \in [k]$ and almost every $x \in \Omega$, and $\Phi$ is indeed a valid strategy profile for the given Blotto or Lotto game, then $\Phi$ is a mixed Nash equilibrium of the game.*

*Proof.* Let $\Phi$ be a mixed strategy profile satisfying the conditions of the theorem. Let $\mathcal{X} \subseteq \Omega$ be the set of points in the battleground for which all the players satisfy the marginal constraint in the theorem. We know from the hypothesis that $\upsilon(\mathcal{X}) = \upsilon(\Omega) = \Upsilon$ and $\beta(\mathcal{X}) = \beta(\Omega) = 1$.

We will show that if players in the set $[k-1]$ fix their strategies, then Player $k$ gains no expected utility by changing his strategy. To this end, let $\Phi_{max} \triangleq \max\limits_{i \in [k-1]} \Phi_i$ be the random function that Player $k$ is effectively competing against, and $M$ be the CDF of $\Phi_{max}(x)$ for almost every $x$. In the remainder of the discussion, for a given random variable $R$, we will use $F_R(x)$ to denote its CDF. Notice that by hypothesis, $M(x) \sim \text{Unif}[0, \frac{k}{\Upsilon}\frac{d\upsilon}{d\beta}(x)]$, and therefore $F_M(x) = x\frac{\Upsilon}{k}\frac{d\beta}{d\upsilon}$ in the domain $[0, \frac{k}{\Upsilon}\frac{d\upsilon}{d\beta}(x)]$. We now bound the expected utility of Player $k$ if he switches to the strategy $\psi$.

$$\begin{aligned}
\mathbb{E}[U_k] &= \mathbb{E}\int_\Omega \frac{\mathbb{1}\left(\psi \in \arg\max_{f \in \{\phi_i\}_{i \in [k-1]} \cup \{\psi\}}\{f(x)\}\right)}{\left|\arg\max_{f \in \{\phi_i\}_{i \in [k-1]} \cup \{\psi\}}\{f(x)\}\right|} d\upsilon(x) \\
&\leq \mathbb{E}\int_\Omega \mathbb{1}\left(\psi \in \text{argmax}_{f \in \{\phi_i\}_{i \in [k-1]} \cup \{\psi\}}\{f(x)\}\right) d\upsilon(x) \\
&= \mathbb{E}\int_\Omega \mathbb{1}\left(M(x) \leq \psi(x)\right) d\upsilon(x) \\
&= \int_\mathcal{X} \mathbb{E}[\mathbb{1}(M(x) \leq \psi(x))] d\upsilon(x) \\
&= \int_\mathcal{X} F_M(\psi(x)) d\upsilon(x) \\
&\leq \int_\mathcal{X} \psi(x) \frac{\Upsilon}{k}\frac{d\beta}{d\upsilon} d\upsilon = \frac{\Upsilon}{k}\int_\mathcal{X} \psi(x) d\beta(x) \leq \frac{\Upsilon}{k}\mathcal{B}_k
\end{aligned}$$

In fact, the calculation above shows that even if Player $k$ switches to a randomized strategy $\Psi$ satisfying the weaker condition of Lotto, his expected utility will be at most $\frac{\Upsilon}{k}\mathcal{B}_k = \frac{\Upsilon}{k}$. However, by symmetry, we know that his expected utility at the mixed strategy profile $\Phi$ is already $\frac{\Upsilon}{k}$. So, he has no incentive to switch his strategy.

Since all players are symmetric, the above argument applies to each Player $i$, not just Player $k$, and thereby concludes the proof that $\Phi$ must be a mixed Nash Equilibrium. $\square$

Lemma 4.3 gives sufficient conditions for a strategy profile to be in mixed Nash equilibrium for a Measure Space Blotto game. In the remainder of this section, we demonstrate the existence of equilibrium strategy profiles $\Phi$ satisfying the sufficient conditions of Lemma 4.3 for a wide class of Measure Space Blotto games. Interestingly, the proof is by algorithmic construction. In particular, we construct an efficient $O(k)$ time algorithm to sample each player's strategy in such an equilibrium profile. The algorithm requires that we be able to define a function that partitions the domain $\Omega$ into parts of equal value measure. To this end, we say a function $\pi : \Omega \to [k]$ is a *k-equipartition of* $\Omega$ *with respect to measure* $\mu$ if, for each $i \in [k]$, $\mu(\pi^{-1}(i)) = \frac{1}{k}\mu(\Omega)$.

**Theorem 4.4.** *Consider the Blotto game* $\left(k, \Omega, \mathbb{R}_+, \vec{\mathbb{1}}, \beta, \upsilon\right)$, *and let* $\pi : \Omega \to [k]$ *be a k-equipartition of* $\Omega$ *with respect to the value measure* $\upsilon$. *Then, if each of the players independently runs Algorithm 1, the players will be in mixed Nash equilibrium. Furthermore, Algorithm 1 runs in* $O(k)$ *time.*

**Algorithm 1:** Algorithm to sample from a symmetric mixed equilibrium of an Measure Space Blotto game given a *k*-equiparition with respect to the value measure.

**Input:** Measure Space Blotto game $\left(k, \Omega, \mathbb{R}_+, \vec{\mathbb{1}}, \beta, \upsilon\right)$, and a *k*-equipartition of $\Omega$ with respect to the value measure $\pi : \Omega \to [k]$.

**Output:** an implicit representation of function $\phi$ sampled from a symmetric mixed equilibrium strategy for a single player.

**1** Draw $(X_1, \ldots, X_k) \sim \text{Dir}(1/(k-1), \ldots, 1/(k-1))$.
**2** $\phi(x) \leftarrow \frac{kd\upsilon}{\Upsilon d\beta}(x) \cdot X_{\pi(x)}$, for all $x \in \Omega$.
**3 return** $\phi$

*Proof. Correctness*: Assume each Player $i$ gets his strategy $\phi_i$ from an independent run of Algorithm 1. Let $\Phi_i$ be the distribution of outputs that Player $i$ could have received by running the algorithm. We will show that each $\phi_i$ indeed satisfies the budget constraint and $\Phi_i$ indeed satisfies the conditions of Lemma 4.3 and thereby the players will be in equilibrium if they each play an output from independent runs of Algorithm 1. Now, we prove the two claims and algorithmic running time.

(a) Budget constraint: For each $i \in [k]$, let $\Omega_i \triangleq \pi^{-1}(i)$. By the partition of $\pi$, we know that the $\Omega_i$s are disjoint and their union is the full space $\Omega$.

$$\int_\Omega \phi(x) d\beta(x) = \sum_{i=1}^k \int_{\Omega_i} \phi(x) d\beta(x)$$
$$= \sum_{i=1}^k \int_{\Omega_i} \frac{kd\upsilon}{\Upsilon d\beta}(x) \cdot X_i \, d\beta(x)$$
$$= \sum_{i=1}^k X_i \cdot \frac{k}{\Upsilon} \cdot \int_{\Omega_i} d\upsilon$$
$$= \sum_{i=1}^k X_i = 1$$

Where the final equality follows from Definition 4.2(ii).

(b) Marginal constraint: $\Phi(x) \sim \frac{k}{\Upsilon} \cdot \frac{d\upsilon}{d\beta}(x) \cdot \text{Beta}\left(\frac{1}{k-1}, 1\right)$ by Definition 4.2(i).

(c) Running Time: we can sample the Dirichlet variable in $O(k)$ time, using the method of [2] to sample $k$ i.i.d variables $Y_i \sim \texttt{Gamma}\left(\frac{1}{k-1}, 1\right)$ and letting $X_i = \frac{Y_i}{\sum_{\ell=1}^k Y_\ell}$ for all $i \in [k]$.

□

Below we demonstrate the generality and applicability of Theorem 4.4 and Algorithm 1.

**Example 4.5.** *Algorithm 1 works whenever the measure space is value k-equipartitionable. Here are several examples of such Blotto problems with such spaces.*

**Interval Blotto:** *We can define the k-equipartition function $\pi(x)$ to output one plus the integer part of $kx$ for each $x \in [0, 1)$. Notice that any function $\phi$ produced by Algorithm 1 will be a step function with at most $k$ steps.*

**Blotto on the Sphere:** *Here $\Omega = \mathbb{S}^2$—the surface of the standard unit-sphere centered at the origin—and $\beta$ and $\upsilon$ are the uniform Lebesgue measure. We can divide the sphere by $k$ equally spaced longitudes, and define the manifolds between successive longitudes to be the $k$ equal partitions for function $\pi$.*

**War of Ideologies:** *We simply define any partition function $\pi$ that splits the country $\mathcal{M}$ into $k$-equal population submanifolds.*

**Symmetric Homogeneous Blotto on $n$ Battlefields [10]:** *We consider $\Omega = [n]$, and let $\upsilon = \beta$ be the counting measure. Whenever $n \bmod k = 0$, we can define the $k$-equipartition function $\pi : [n] \to [k]$ as $\pi(j) = (j \bmod k) + 1$. This argument is by Boix-Adserà et al.*

## 5 Characterizing the Symmetric Fair Equilibria of Interval Blotto

In this section, we focus our attention on Interval Blotto (defined in Section 3.1). Notice that Algorithm 1 samples a symmetric fair equilibrium for this game, i.e., the marginal distribution $\Phi_i(x)$ is identical for all Players $i$ and all points $x$ in the interval. In particular, the CDF of these marginals is $\mathcal{G} \equiv k \cdot \text{Beta}\left(1/(k-1), 1\right)$. There is no unique symmetric fair equilibrium of the Interval Blotto game. To see this, note that Algorithm 1 outputs a sample from a *different* strategy for every *different* $k$-partition, $\pi$, of the interval that is fed as input. However, the marginals of all these various strategies are also distributed as $\mathcal{G}$. This leads to a natural question: is $\mathcal{G}$ the unique marginal of all symmetric fair equilibria of Interval Blotto? This section is devoted to proving that the answer to this question is "yes".

In our proof, we consider a symmetric fair mixed strategy profile $\Phi = (\Phi_i)_{i=1}^{k}$ for the $k$ players. Since the strategy profile is symmetric and fair, we can define $G$ as the marginal distribution of $\Phi_i(x)$ for each $i \in \{1, \ldots, k\}$ and $x \in [0, 1]$. Our goal is to show that $G \equiv \mathcal{G}$, whenever $\Phi$ is an equilibrium. In our analysis, we will be interested in computing Player $k$'s utility when he plays a particular strategy $\psi_k$ in response to the other players fixing their mixed strategies to be $(\Phi_i)_{i=1}^{k-1}$. So, it is useful to define $\Phi_{max} \triangleq \max_{i \in [k-1]} \Phi_i$, since Player $k$ only gets utility at points of the battleground where $\psi_k$ exceeds $\Phi_{max}$. Noticing that $\Phi_{max}(x)$ has CDF $G^{k-1}$ (since each $\Phi_i$ has CDF $G$), we denote this CDF by $M = G^{k-1}$.

### 5.1 An Approach to Characterization and a Hurdle

In order to show that $G = \mathcal{G}$ is the unique marginal of symmetric fair equilibria of Interval Blotto, we need to demonstrate that whenever Players $1, \ldots, k-1$ choose to play according to some strategy profile $(\Phi_i)_{i=1}^{k-1}$ with marginal $G \neq \mathcal{G}$, Player $k$ can *exploit* that strategy profile to play some other randomized strategy $\Psi_k$ that wins him excess utility. A natural idea to construct such a $\Psi_k$ if we were considering a set of $n$ discrete battlefields (rather than the interval) would be to start with the marginal CDF $G$, manipulate it to produce some other CDF $G'$ (with the same expected value as $G$), and show that if Player $k$ were to play $G'$ at each of the $n$ battlefields, he would win more than $1/k$ of the utility.

The above idea is appealing, because it focuses on analyzing the single marginal distribution $G$ rather than analyzing the entire distribution over measurable functions $\Phi_{max}$. However, it seems to fatally require the battleground to be composed of a discrete set of battlefields—a condition $[0, 1]$ does not satisfy—and does not specify a way to ensure the coupleability of the resultant $n$ marginals of distribution $G'$ into a joint distribution that satisfies the hard budget constraint of Blotto. In particular, even if we are able to construct a marginal distribution $G'$ with expected value 1 that exploits $G$, it would take complex measure theoretic arguments to show that there exists a distribution $\Psi_k$, over functions integrating to 1 (i.e. satisfying the hard budget constraint), that has all marginals distributed as $G'$.

We overcome this hurdle by inverting CDFs. In particular, assuming that all distributions have strictly monotonic CDFs, we can informally state our observation as follows. Given that Players $1, \ldots, k-1$ have fixed their symmetric fair randomized strategies and their randomness is independent of Player $k$'s, Player $k$ gains the same expected utility by playing the deterministic strategy $\psi_k = (G')^{-1} : [0,1] \to \mathbb{R}_+$, as he would by playing the hypothesized strategy $\Psi_k$ that has all marginals distributed as $G'$. Our observation, which we prove as Lemma 5.3 below, relies on the linearity of expectation, and the fact that a variable with CDF $G'$ is distributed by the pushforward measure of the uniform distribution on $[0,1]$ by the function $(G')^{-1}$.

Formally, a CDF only has a true inverse when it is a strictly monotonic function, and in general a CDF's inverse is not unique. We eliminate this concern by defining the following unique inverse.

**Definition 5.1.** Let $H(x)$ be a cumulative distribution function (CDF) for a random variable $X$ that has the support $\mathcal{S}$. We define $H^{-1}(p) \triangleq \arg\sup_{\theta}\{H(\theta) \leq p\}$ for every $p \in [0,1]$. If $H$ is strictly monotonic and continuous, then our definition corresponds to the standard inverse, while in other cases, our definition smooths out jumps in the distribution and resolves points that have multiple inverses. △

Notice that the inverse-CDF has the following nice property.

**Lemma 5.2.** *The inverse-CDF $H^{-1}$ has the property $\int_{[0,1]} H^{-1}(x)dx = \mathbb{E}_{X \sim H}[X]$*

*Proof.* Graphically, the lemma is proved by seeing that both expressions denote the area between the $y$-axis and the curve of $H(x)$. Algebraically, we perform a change of variables using: $x = H(y)$, whereby $y = H^{-1}(x)$ and $dx = dH(y)$. We see that:

$$\int_{[0,1]} H^{-1}(x)dx = \int_{\text{supp}(H)} y\,dH(y) = \mathbb{E}_{X \sim H}[X].$$

□

Finally, we formalize our observation from above in this key lemma.

**Lemma 5.3.** *Let $H$ be a CDF and $M$ be a distribution with no atoms. If $\mathbb{E}_{X \sim H}[X] = 1$, then Player $k$'s expected utility for defecting to the strategy $H^{-1}$ when all other Players $i \in [k-1]$, play $\Phi_i$ is $\mathbb{E}_{X \sim H} M(X) = \mathbb{P}_{X \sim H, Y \sim M}(Y \leq X)$.*

*Proof.* The proof is a chain of inequalities, where the main operation is a change of variables using: $x = H(y)$, whereby $y = H^{-1}(x)$ and $dx = dH(y)$. Let $U_k$ be the utility obtained by Player $k$ when he plays $H^{-1}$, we calculate:

$$\mathbb{E}[U_k] = \mathbb{E}\left[\int_{[0,1]} \mathbb{1}(\Phi_{max}(x) \leq H^{-1}(x))dx\right]$$

$$= \int_{[0,1]} \mathbb{P}(\Phi_{max}(x) \leq H^{-1}(x))dx$$

$$= \int_{[0,1]} M(H^{-1}(x))dx$$

$$= \int_{supp(H)} M(y)dH(y)$$

$$= \mathbb{E}_{X \sim H}[M(X)]$$

$$= \mathbb{P}_{X \sim H, Y \sim M}(Y \leq X)$$

□

## 5.2 Uniqueness Proof

We are now ready to show that the marginals of the equilibrium from Theorem 4.4 are in fact characteristic of all symmetric fair equilibria of Interval Blotto, i.e. that $G = \mathcal{G}$ for all symmetric fair equilibria $\Phi$. The first part of our proof analyzes randomized strategies to eliminate the possibilities of atoms in the marginal $G$, then we use the inverse-CDF lemma in conjunction with some analytic techniques to complete the proof.

**Theorem 5.4.** *Let $\Phi = (\Phi_i)_{i=1}^k$ be a symmetric fair mixed Nash equilibrium of Interval Blotto, and let $G$ be the CDF of the unique marginal distribution, i.e., $\Phi_i(x) \sim G$ for all $i \in [k]$ and $x \in [0, 1]$. Then, $G \sim k \cdot \text{Beta}\left(\frac{1}{k-1}, 1\right)$.*

*Proof.* Let $\mathcal{S}$ be the support of $G$. Our proof strategy is to successively restrict the possible forms of $G$ using the fact that $\Phi$ is in equilibrium until we are left with only the Beta distribution. In tandem with Theorem 4.4, this will show that the Beta distribution $\mathcal{G}$ is the unique marginal distribution of all symmetric fair equilibria of Interval Blotto.

To that end, let Players 1 through $k - 1$ fix their strategies $\Phi_1, \ldots, \Phi_{k-1}$, and Observe that by symmetry, Player $k$ wins $1/k$ expected utility by sticking to the prescribed strategy $\Phi_k$. We will now show many properties of $G$.

1. $\mathbb{E}_{Z \sim G}[Z] = 1$: the proof simply follows from the fact that $\Phi_1$ is fair, and that any realization, $\phi_i$, of it is a function that integrates to 1. Alternatively:

$$\mathbb{E}\left[\int_0^1 \Phi_1(x)dx\right] = 1 \implies \int_0^1 \mathbb{E}_{Z \sim G}[Z]dx = 1$$
$$\implies \mathbb{E}_{Z \sim G}[Z] = 1$$

2. *G has no atoms:* Assume to the contrary that $G$ has an atom of weight $\eta > 0$ at $a \in \mathcal{S}$, and that this is an atom of largest weight. For each $\delta \in (0, 1)$, define the distribution $\Psi_\delta$ over measurable functions $\psi : [0, 1] \to \mathbb{R}_+$ with integral equal to 1 as follows:

   (a) Sample $\psi \sim \Phi_k$
   (b) Set $\psi(x) \leftarrow 0$ for all $x \in [0, \delta)$

(c) Set $\psi(x) \leftarrow a + \epsilon$ wherever $\psi(x) = a$ for very small $\epsilon$ so that $\psi$ integrates to at most 1

(d) If $\psi$ integrates to less than 1, increase it arbitrarily at points so it remains measurable but integrates to exactly 1

(e) Output $\psi$

We will show that Player $k$ will gain an advantage by defecting to $\Psi_\delta$ for some (small) $\delta > 0$. Let $\mathcal{X}_a = \bigcap_{i=1}^{k} \Phi_i^{-1}(a)$ be the set of points where all the players play exactly $a$. Since $G$ has an atom of mass $\eta$ at $a$, the expected measure $\mathbb{E}[\lambda(\mathcal{X}_a)] = \eta^k$ because

$$\mathbb{E}[\lambda(\mathcal{X}_a)] = \mathbb{E}\left[\int_{[0,1]} \mathbb{1}\{x \mid \forall i, \Phi_i(x) = a\}\right]$$

$$= \int_{[0,1]} \mathbb{P}\left(\bigwedge_{i=1}^{k} \Phi_i(x) = a\right) dx$$

$$= \int_{[0,1]} \eta^k dx = \eta^k \cdot \lambda([0,1]) = \eta^k$$

Deviating from $\Phi_k$ to $\Psi_\delta$ makes Player $k$ gain utility on $\mathcal{X}_a \cap [\delta, 1]$ and lose utility on $[0, \delta)$. Quantitatively, the total gain is at least $\frac{k-1}{k}(1-\delta) \cdot \eta^k - \frac{1}{k}\delta$ by the above calculation. Since $\eta$ is a fixed constant, this gain is larger than 0 whenever $\delta$ is sufficiently small. Thus, for some $\delta^* \in (0, 1)$, Player $k$ gains by deviating to $\Psi_{\delta^*}$; a contradiction. Thus, $G$ has no atoms.

3. *G is strictly increasing within its support:* Assume to the contrary that $0 \leq G(a) = G(b) < 1$ for $b > a + \eta$, i.e., that the CDF $G$ is constant between $a$ and $b$ that are at least $\eta$ far apart. Further, ensure that for any $\Delta > 0, G(b + \Delta) > G(b)$. $G$ is monotonic and continuously non-decreasing in the interval $[b, b + \Delta]$ for any $\Delta > 0$ that is sufficiently small; choose any $\Delta < \eta$.

For every $\delta \in (0, \Delta)$, define $H_\delta$ to be a distribution that modifies $G$ as follows:

(a) Let $\mu = G(b + \delta) - G(b)$

(b) $H_\delta$ moves $\mu/2$ mass in $G$ from $[b, b + \delta)$ to an atom at $b + \Delta$

(c) $H_\delta$ moves the remaining $\mu/2$ mass in $G$ from $[b, b+\delta)$ to an atom at $a + \epsilon$ where $\epsilon > 0$ is picked precisely to ensure that $H_\delta$ has expectation equal to 1 just like $G$. (This is indeed possible because $\Delta \leq \eta$.)

Now, define $\psi_\delta = H_\delta^{-1}$, this a legal deterministic strategy for Player $k$ by Lemma 5.2. The expected utility gained by Player $k$ for defecting to this strategy is $\mathbb{P}_{X \sim H_\delta, Y \sim M}(Y \leq X) - \mathbb{P}_{Z \sim G, Y \sim M}(Y \leq Z)$ by Lemma 5.3. Notice that whenever $\mathbb{P}_G([b+\delta, b+\Delta)) > \mathbb{P}_G([b, b+\delta)) > 0$, this gain is positive, since the $\mu/2$ mass that was moved up causes more gains than the loss due to the $\mu/2$ mass that was moved down. Since $G$ is continuous in the interval from $b$ to $b + \Delta$, we can choose $\delta^* > 0$ sufficiently small such that this criterion is met, which is a contradiction. Therefore, we know that the $G$ is strictly increasing in its support. That is, $\mathcal{S} = [0, t]$, $G(0) = 0$, $G(1) = 1$, and $G$ is strictly increasing from 0 to 1 in the interval $[0, t]$.

4. $M = G^{k-1}$ *is linear in its support* $[0, t]$: Note that $G^{-1}$ is a continuous function of integral 1 by Lemma 5.2, and if Player $k$ defects to $\psi = G^{-1}$ he gets an expected utility of $1/k$ by Lemma 5.3. Now, consider any values $0 \leq a < b \leq t$, and small $\epsilon > 0$. We define a deterministic strategy $\psi_{a,b,\epsilon}$ using these parameters as follows:

(a) Define $\mathcal{X}_{a,\epsilon} = G^{-1}((a-\epsilon, a+\epsilon)) \subseteq [0,1]$.

(b) Define $\psi_{a,b,\epsilon}$ to be $G^{-1}$ but modified in the interval $\mathcal{X}_{a,\epsilon}$ to be a step function taking values $0$ and $b$ such that $\psi_{a,b,\epsilon}$ integrates to $1$.

Defecting to $\psi_{a,b,\epsilon}$ modifies his expected utility calculation in the interval $\mathcal{X}_{a,\epsilon}$. If he plays $G^{-1}$, he gets an expected utility of

$$\mathbb{E}\left[\int_{\mathcal{X}_{a,\epsilon}} \mathbb{1}(\Phi_{max}(x) \leq G^{-1}(x))dx\right] = \int_{\mathcal{X}_{a,\epsilon}} M(G^{-1}(x))dx$$
$$= \lambda(\mathcal{X}_{a,\epsilon})(M(a) + \epsilon')$$

where $\epsilon' \in [-\epsilon, \epsilon]$ by the mean value theorem. On the other hand, if he plays $\psi_{a,b,\epsilon}$ he gets an expected utility of

$$\mathbb{E}\left[\int_{\mathcal{X}_{a,\epsilon}} \mathbb{1}(\Phi_{max}(x) \leq \psi_{a,b,\epsilon}(x))dx\right] = \frac{a+\epsilon''}{b}\lambda(\mathcal{X}_{a,\epsilon})M(b)$$

where $\epsilon'' \to 0$ as $\epsilon \to 0$ Since $\psi_{a,b,\epsilon} \to G^{-1}$ as $\epsilon \to 0$, we see that $M(a) = \frac{a}{b}M(b)$. This implies that $\forall a,b \in [0,1], M(a)/a = M(b)/b$.

5. $M(x) = x/k$ in the range $[0, k]$: First, observe that $t > 1$. Otherwise, Player $k$ could defect to the constant function $\psi(x) = 1$ and win the whole interval. The constant function $M(x)/x$ must be $1/k$ to account for the fact that playing the function $\psi(x) = 1$ gets Player $k$ a utility of $1/k$. This implies that $M(x) = x/k$ in the range $[0, t]$. Since we additionally know that $\mathbb{E}_{Z \sim G}[Z] = 1$ and $G^{k-1} = M$, we can specify $t$ uniquely to be $k$ and $G$ to be the CDF of $k \cdot \text{Beta}\left(\frac{1}{k-1}, 1\right)$.

Because $M(x) = x/k$, we conclude that $G$ must be distributed as $k \cdot \text{Beta}\left(\frac{1}{k-1}, 1\right)$. □

## 6 Remarks & Open Problems

In this paper, we have extended the definition of the Colonel Blotto game to model multiplayer competitions over arbitrary measure spaces, and analyzed the symmetric equilibria of such Measure Space Blotto games. Our generalized definition captures all the old applications of Blotto, such as multiparty elections, and also opens the door to analyzing competitions over continuous manifolds such as the Competition of Data Providers (Section 1.2).

Our main algorithmic result is a linear time algorithm to sample an equilibrium strategy of any symmetric $k$ player Measure Space Blotto game played on a $k$-equipartitionable measure space (Algorithm 1). This result subsumes Boix-Adsera et al.'s algorithm for standard Blotto with more than three players—the only equilibrium result known previously for any type of Blotto game with more than three players [10]. A technical highlight of the paper is the characterization of symmetric fair equilibria of Interval Blotto (Theorem 5.4). This result was proved by exploiting the continuity of the battleground to map from marginal distributions to inverse-CDF allocation strategies.

Computing efficiently sampleable equilibria of multiplayer asymmetric Blotto games is a tantalizing open problem for both the standard and measure-space settings. From a characterization stand point, we conjecture that all symmetric equilibria of Interval Blotto are also fair; and are curious to see research that confirms or refutes this hypothesis. Finally, we look forward to seeing applied work in economics and the social sciences that uses the Generalized Blotto framework in this paper to understand strategic scenarios in the real world.

# 7 Acknowledgements

I would like to thank Enric Boix-Adserà, Costis Daskalakis, Ben Edelman, Sohil Shah, Sarath Pattathil, and Yury Polyanskiy for helpful discussions. I would like to thank Ben Edelman in particular for carefully reading through the paper and giving invaluable comments.

# References

[1] AmirMahdi Ahmadinejad, Sina Dehghani, MohammadTaghi Hajiaghayi, Brendan Lucier, Hamid Mahini, and Saeed Seddighin. From duels to battlefields: Computing equilibria of blotto and other games. *Mathematics of Operations Research*, 44(4):1304–1325, 2019.

[2] Joachim H Ahrens and Ulrich Dieter. Computer methods for sampling from gamma, beta, poisson and bionomial distributions. *Computing*, 12(3):223–246, 1974.

[3] Daniel G Arce, Rachel TA Croson, and Catherine C Eckel. Terrorism experiments. *Journal of Peace Research*, 48(3):373–382, 2011.

[4] Michael R Baye, Dan Kovenock, and Casper G De Vries. The all-pay auction with complete information. *Economic Theory*, 8(2):291–305, 1996.

[5] Soheil Behnezhad, Avrim Blum, Mahsa Derakhshan, MohammadTaghi HajiAghayi, Mohammad Mahdian, Christos H Papadimitriou, Ronald L Rivest, Saeed Seddighin, and Philip B Stark. From battlefields to elections: Winning strategies of blotto and auditing games. In *Proceedings of the Twenty-Ninth Annual ACM-SIAM Symposium on Discrete Algorithms*, pages 2291–2310, Philadelphia, PA, 2018. SIAM.

[6] Soheil Behnezhad, Avrim Blum, Mahsa Derakhshan, Mohammadtaghi Hajiaghayi, Christos H. Papadimitriou, and Saeed Seddighin. Optimal strategies of blotto games: Beyond convexity. In *Proceedings of the 2019 ACM Conference on Economics and Computation*, EC '19, page 597–616, New York, NY, 2019. Association for Computing Machinery.

[7] Soheil Behnezhad, Sina Dehghani, Mahsa Derakhshan, MohammadTaghi HajiAghayi, and Saeed Seddighin. Faster and simpler algorithm for optimal strategies of blotto game. In *Thirty-First AAAI Conference on Artificial Intelligence*, pages 369–375, Palo Alto, California, 2017. AAAI.

[8] Robert M. Bell and Thomas M. Cover. Competitive optimality of logarithmic investment. *Mathematics of Operations Research*, 5(2):161–166, 1980.

[9] Donald W Blackett. Pure strategy solutions of blotto games. *Naval Research Logistics Quarterly*, 5(2):107–109, 1958.

[10] Enric Boix-Adserà, Benjamin L. Edelman, and Siddhartha Jayanti. The multiplayer colonel blotto game, 2020.

[11] Emile Borel. The theory of play and integral equations with skew symmetric kernels. *Econometrica*, 21(1):97–100, 1953. orig. published 1921.

[12] Emile Borel and Jean Ville. *Application de la théorie des probabilités aux jeux de hasard*. Gauthier-Vilars, Paris, 1938.

[13] Pern Hui Chia and John Chuang. Colonel blotto in the phishing war. In *Decision and Game Theory for Security*, pages 201–218, Berlin, Heidelberg, 2011. Springer.

[14] Emmanuel Dechenaux, Dan Kovenock, and Roman M Sheremeta. A survey of experimental research on contests, all-pay auctions and tournaments. *Experimental Economics*, 18(4):609–669, 2015.

[15] Marcin Dziubiński. Non-symmetric discrete general lotto games. *International Journal of Game Theory*, 42(4):801–833, 2013.

[16] Lawrence Friedman. Game-theory models in the allocation of advertising expenditures. *Operations Research*, 6(5):699–709, 1958.

[17] Russell Golman and Scott E Page. General blotto: games of allocative strategic mismatch. *Public Choice*, 138(3-4):279–299, 2009.

[18] Oliver Gross and Robert Wagner. A continuous colonel blotto game. Technical Report RM-408, RAND Corporation, 1950.

[19] Oliver Alfred Gross. The symmetric blotto game. Technical Report RM-424, RAND Corporation, 1950.

[20] Sergiu Hart. Discrete colonel blotto and general lotto games. *International Journal of Game Theory*, 36(3-4):441–460, 2008.

[21] Rafael Hortala-Vallve and Aniol Llorente-Saguer. Pure strategy nash equilibria in non-zero sum colonel blotto games. *International Journal of Game Theory*, 41(2):331–343, 2012.

[22] Dan Kovenock and Brian Roberson. Generalizations of the general lotto and colonel blotto games. Working Papers 15-07, Chapman University, Economic Science Institute, 2015.

[23] Dmitriy Kvasov. Contests with limited resources. *Journal of Economic Theory*, 136(1):738–748, 2007.

[24] Jean-François Laslier. How two-party competition treats minorities. *Review of Economic Design*, 7(3):297–307, 2002.

[25] Jean-François Laslier and Nathalie Picard. Distributive politics and electoral competition. *Journal of Economic Theory*, 103(1):106 – 130, 2002.

[26] Jennifer Merolla, Michael Munger, and Michael Tofias. In play: A commentary on strategies in the 2004 us presidential election. *Public Choice*, 123(1-2):19–37, 2005.

[27] Roger B. Myerson. Incentives to cultivate favored minorities under alternative electoral systems. *American Political Science Review*, 87(4):856–869, 1993.

[28] John F. Nash. Equilibrium points in n-person games. *Proceedings of the National Academy of Sciences*, 36(1):48–49, 1950.

[29] J. Naskar and A. C. Maioli. Quantum multiplayer colonel blotto game, 2020.

[30] Joydeep Naskar. Comments on quantization of colonel blotto game. *Preprint*, 2020.

[31] Brian Roberson. The colonel blotto game. *Economic Theory*, 29(1):1–24, 2006.

[32] Brian Roberson and Dmitriy Kvasov. The non-constant-sum colonel blotto game. *Economic Theory*, 51(2):397–433, 2012.

[33] Galina Schwartz, Patrick Loiseau, and Shankar S Sastry. The heterogeneous colonel blotto game. In *2014 7th International Conference on NETwork Games, COntrol and OPtimization (NetGCoop)*, pages 232–238, New York, NY, 2014. IEEE.

[34] Martin Shubik and Robert James Weber. Systems defense games: Colonel blotto, command and control. *Naval Research Logistics Quarterly*, 28(2):281–287, 1981.

[35] Caroline Thomas. N-dimensional blotto game with heterogeneous battlefield values. *Economic Theory*, 65(3):509–544, 2018.

[36] John W. Tukey. A problem of strategy. *Econometrica*, 17(1):73, 1949.

[37] Jonathan Weinstein. Two notes on the blotto game. *The BE Journal of Theoretical Economics*, 12(1):1–11, 2012.